# Unix Memory Allocations Are Not Poisson

James Garnett *Student Member, IEEE,* and Elizabeth Bradley *Member, IEEE*

*Abstract*— **In multitasking operating systems, requests for free memory are traditionally modeled as a stochastic counting process with independent, exponentially-distributed interarrival times because of the analytic simplicity such *Poisson* models afford. We analyze the distribution of several million unix page commits to show that although this approach could be valid over relatively long timespans, the behavior of the arrival process over shorter periods is decidedly not Poisson. We find that this result holds regardless of the originator of the request: unlike network packets, there is little difference between system- and user-level page-request distributions. We believe this to be due to the bursty nature of page allocations, which tend to occur in either small or extremely large increments. Burstiness and persistent variance have recently been found in *self-similar* processes in computer networks, but we show that although page commits are both bursty and possess high variance over long timescales, they are probably not self-similar. These results suggest that altogether different models are needed for fine-grained analysis of memory systems, an important consideration not only for understanding behavior but also for the design of online control systems.**

## I. Introduction

TRAFFIC studies are investigations of the pattern of arrivals of service requests to a computing system. Knowing these patterns is critical to understanding these systems, since the manner in which requests are satisfied determines the fashion in which local (and sometimes even remote, downstream) resources are consumed. This kind of knowledge allows system engineers to predict and accomodate potential bottlenecks. In a communications network, traffic consists of packets moving between network nodes. In the past, it was standard practice to assume that packet arrivals were uncorrelated but identically distributed, in much the same manner as individual atomic decays in radioactive material. At a large enough timescale, this kind of behavior creates a *Poisson* rate of stochastic arrivals. Poisson arrivals are easy to work with for a variety of reasons, ranging from the simplicity of mathematical modeling to predictability of average resource consumption and the expected variance from that average. Recent studies of network traffic, however, have shown that traditional assumptions of Poisson arrival behavior are for the most part invalid, and that network data arrival processes appear to possess *self-similar* properties [19], [25], [30] instead. Such processes exhibit bursts of highly correlated arrivals with no natural period and high variance over long timescales. The difference between the Poisson and self-similar arrivals is immediately apparent on graphs of their temporal distribution. Figure 1 compares the arrival rate of a known self-similar process (the BC-pAug89 LAN packet trace from the Internet Traffic Archive [1], [19]) to a Poisson process with the same mean interarrival time. It is apparent that the self-similar packet trace of Figure 1 is not "smooth" at any timescale (in contrast to the Poisson data), but instead exhibits high, persistent, long-term variance. This is a consequence of the *scaling* of self-similar processes, which we describe formally as follows. A process $X = \{X(t), t \in \mathbb{R}\}$ is *self-similar with index H* if, for any $a > 0$, the finite-dimensional distributions of $\{X(at), t \in \mathbb{R}\}$ are the same as those of $\{a^H X(t), t \in \mathbb{R}\}$ [28]. Thus, as a self-similar process is scaled in time, its *distribution* (and associated variance) remains the same as the original.

Formal characterization of arrival processes is becoming important as research interest turns towards implementing principled control systems for resource management [15]. The control goal in this application is to allocate resources such that the system operates efficiently under heavy user task loads, with the least amount of resources taken up by the management system. Service demands—which can be modeled as a stochastic arrival process—directly affect resource levels, so understanding the arrivals is the key to understanding resource usage patterns and to designing controllers that optimize those patterns. Current control methods are not this sophisticated, instead emphasizing simple heuristic methods to manage the underlying resources—methods that can lead to extremely inefficient operation. Even in modern virtual memory systems, the customary "control" paradigm is simply to allocate as many pages as the running software requests, up to the point at which no semiconductor memory remains unallocated. Beyond this point, the classic response is to begin *paging*: swapping out pages between disk (virtual memory) and RAM (semiconductor memory), and hopefully not evicting RAM-based pages of running processes. Such strategies can create a tremendous load on a system, and so a large body of research exists on how to page, e.g. [2], [3], [5], [9], [10], [17], [22], [26]. An alternative is to determine how to minimize paging by the prudent application of smarter control schemes, using predictions of demand to determine which threads are granted memory, and when. This approach is not unheard of: when passive TCP/IP congestion control schemes proved inadequate to the task of handling the increasing traffic load on the Internet, for instance, the IETF (Internet Engineering Task Force) recommended a policy of Active Queue Management: using active controllers to reduce network congestion. The first generation of these were designed using Poisson packet-arrival assumptions and hence were ineffective (such as [21]). Present research efforts focus on crafting controllers for particular traffic patterns [11], [12]. These new approaches rely upon extensive network traffic studies that have conclusively shown that such traffic is self-similar, rather than Poisson [13], [19], [25], [30].

The first step towards building smarter controllers for memory systems is to study the arrival pattern of requests for memory pages. Page commits in these systems are also typically modeled as Poisson processes (as in [6]), but no studies of the actual nature of the arrivals have been performed. To explore this, we instrumented unix kernels on machines in two very different environments on a departmental network in order to record page allocations. We found, in general, that even over timescales on the order of seconds, the page-request arrival process is not realistically modeled by Poisson models—but probably not by self-similar processes, either. This appears to be due to the structure of the bursts, which are created by underlying software processes that are unrelated to the virtual memory system itself, and that tend to occur in either small or extremely large increments, as discussed in Section 3.

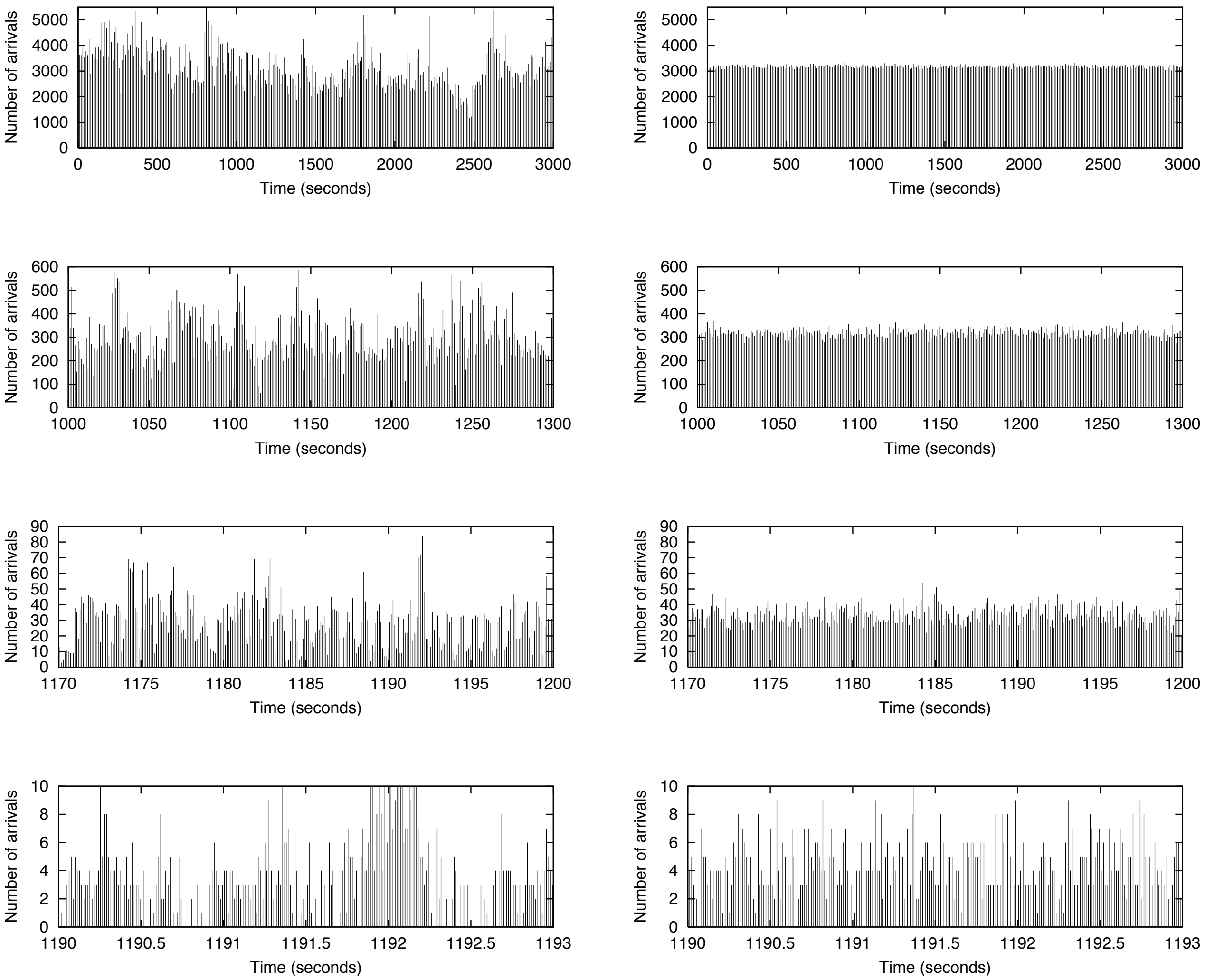

Fig. 1. Self-similar and Poisson time-series data. From top to bottom are 3000, 300, 30 and 3 seconds of actual network traffic arrivals (left) and those from a Poisson simulation (right) with the same mean arrival rate. Self-similar traffic variance is higher and more persistent than its Poisson counterpart, which converges towards a low-variance mean at long timescales.

For our study, we instrumented the OpenBSD-2.9 Unix kernel to capture the time history of page commits and to categorize each allocation based upon the type of the requesting process, i.e., whether due to system or user activity. OpenBSD is a derivative of the BSD-4.4 operating system developed at Berkeley; it is in wide use as a firewall OS because of its high performance and security features [27]. OpenBSD uses Cranor's UVM virtual memory system, a recent high performance VM system [7]. It is thus representative of a modern general-purpose, production-quality unix, and its active development and freely available source make it attractive as a research platform. In OpenBSD, once memory for kernel data structures has been allocated at boot time, all future page allocations are made through the `uvm_pagealloc()` routine, often after the occurence of a page fault. We modified this routine to record the successful commits to a ring buffer, distinguishing each one as either:

- a kernel page fault allocation,
- a root-user page fault,
- a nonroot-user pagefault, or
- an allocation made from a nonfaulted state (i.e., a direct call to `uvm_pagealloc()`).

A specially designed program then periodically dumped the kernel ring buffer to a local disk. In Section 2 we describe both this program and the kernel instrumentation in detail, since these two aspects of the study had the potential to affect the results that they were meant to record. Of particular note for that discussion, and for the remainder of the paper, is that we use the term **thread** in this paper to mean *thread of execution*— what is normally referred to as a unix *process*— in order to reserve the term "process" to denote a stochastic process.

We chose two specific OpenBSD systems in a university computer science department for our study, representing environments at opposite ends of the virtual memory performance spectrum: a heavily loaded network server and a relatively idle user workstation. The server acts as the SMTP (email) hub and also handles majordomo and Mailman mailing lists, primary DNS for several Internet domains, NTP (time) queries, and a English-word dictionary server for internal queries. This system has only a simple terminal console—no X11—and experiences no normal user activity except an occasional administrative login. The second is a help-desk machine at the computer operations service window that is used by a series of different persons each day. It runs X11 and experiences primarily user-based activity, e.g. logins to other machines, hostname lookups, class-schedule verifications for students, etc. We refer to the data from the first machine as CU-SERVER, and that from the second as CU-WORKSTATION, to distinguish the primarily server-based and interactive-user natures of the two machines.

We collected data over two independent several-day periods and examined the resulting page commit time-series data on a categorized basis. We collected 10,650,956 distinct page allocations in the CU-SERVER set and 1,978,623 in the CU-WORKSTATION set. To give an idea of the level of activity on the two machines, the CU-SERVER data was collected over an approximately three-day period (Friday, August 24, 2001 through Monday, August 27, 2001) whereas the much smaller CU-WORKSTATION set was collected over a 14-day period (Wednesday, September 5, 2001 through Wednesday, September 19, 2001). As would be expected, the server experienced a great many more page requests. Our goal was to determine whether Poisson models would be consistent not only with the general allocation arrival process, but also with the *subprocesses* from the different kinds of requesting threads. We therefore analyzed the data by examining individual categories of page commits representing primarily *system* activity (kernel, root-user, and nonfaulted memory allocations) and *user* (nonroot-user) allocations. We expected that system memory usage—those allocations made in response to kernel or timer-based threads—would tend to be correlated, and therefore "less Poisson," whereas user allocations would occur more in response to independent (human) decisions and thus might be well-modeled by a Poisson process. Recent work on wide-area network traffic justifies this belief; in [25], for example, arrivals of SYN packets in the initial connection attempts of TELNET or rlogin sessions are shown to be consistent with Poisson arrivals, from which the authors concluded that such arrivals are due to the random event of a human deciding to initiate a login session. Implicit in our approach is the assumption that there are enough arrivals at the timescales of interest to justify saying that a process is Poisson, or not. Poisson or nearly Poisson behavior is the result of enough mostly uncorrelated arrivals for the law of large numbers to prevail, a notion illustrated well by the 'smoothing out' of the Poisson data of Figure 1. We believed that it might be possible, even at the short timescales of page commits, to find Poisson-modelable arrivals. Since threads related to user-level activity were most likely to exhibit Poisson arrivals, we analyzed the page-request data by combining categories of requests in a hierarchy ranging from nearly purely user requests to those created almost entirely by system threads. We attempted to identify Poisson effects and relate them to user activity in order to investigate whether a large conglomerate of processes could result in an overall Poisson pattern of arrivals, even if the correlation between system events were included, which in turn would indicate that those correlations were weak. The designations used for the different types of allocations are given in Table 1. Since most root threads in a system are daemons performing system-level or automated tasks, and since only the kernel makes direct, non-faulted calls to the `uvm_pagealloc()` routine, we felt that these categories adequately represented a reasonable interpretation of "system" versus "user" activity.

TABLE I

PAGE COMMIT DESIGNATIONS

| *Designation* | *Type of allocation* |
|---|---|
| ALL | All page commits combined |
| SYSTEM | Kernel, Root-user and Nonfaulted commits |
| USERFIRST | First commit from a new nonroot-user thread |
| USERPOST | Other commit from nonroot-user threads |

The results are twofold. As we describe in Section 3, Poisson modeling of *any* category of the data, with either fixed- or variable-rate Poisson models, is not justified. We believe this to be a result of the environment of the machines under study; there is a strong relationship between the tasks that a machine performs and its page-request distribution. Our second result, covered in Section 4, is that the page-request data is *heavy tailed*[1] and possesses high variance that persists over long timescales: effects that are both inconsistent with non-Poisson behavior. Heavy tails, in combination with persistently high variance, can be indicative of the long-term burstiness that characterizes self-similarity, a useful result that implies that simply replacing Poisson models with self-similar ones might be effective. However, we find that page commit arrival process does not appear to be strictly self-similar, either. We conclude by discussing some of the practical and theoretical implications of these results for modeling of unix memory systems.

## II. Study Systems and Data Collection

The systems used in this study consisted of an 800Mhz AMD Athlon with 512Mb of memory for the CU-SERVER dataset and a 233Mhz Intel Pentium with 64Mb of memory for the CU-WORKSTATION data, both fairly high-performance machines that have the potential to generate large amounts of trace data. The server, for example, was capable of generating 30,000 page allocations in a single second, each of which needed to be time stamped and recorded. Great care was therefore taken to ensure that our results were not unduly influenced by the data collection itself.

The steps we took to collect the necessary data consisted of (1) modifying the OpenBSD-2.9 kernel to time stamp, type and record each page commit in an in-kernel static ring buffer; and (2) writing a user-level program that dumped the ring buffer to disk at intervals corresponding to the arrivals of an exponential arrival process, as will be described in this section. The kernel buffer consisted of 65,536 *slots*, each of which could hold all the information about one allocation:

- the *time* of the allocation, accurate to 0.01s
- its *type*, indicating whether it resulted from
  - a kernel page fault
  - a root-user page fault
  - a nonroot-user page fault
  - a direct call to `uvm_pagealloc()`

The time of the allocation, recorded at the entry point to `uvm_pagealloc()`, was taken from the kernel **mono_time** variable, a clock variable that gives the time in seconds and microseconds since boot and is guaranteed to be monotonically increasing. Since the distribution of network TCP SYN packets (the first in a stream of packets that synchronizes and starts a TCP connection) has been found to be Poisson [25], we also wished to track the first page commit by each new nonroot-user thread. To do this, we modified the kernel `fork()` system by adding an array of PID_MAX flags, where PID_MAX is the maximum thread ID. Whenever a new thread was forked, its corresponding flag in this array was set, which could then be noted when page allocations from it began to appear. Finally, the kernel instrumentation was enabled with a static flag that could be set or cleared as desired; this allowed us to boot the instrumented kernels, allow

[1]A heavy-tailed distribution is one that decays more slowly than exponentially, so the probability mass is shifted more towards the tail than in an exponential distribution. In a heavy tailed distribution of arrivals, where one plots interarrival intervals versus probability, this would mean that longer intervals would be more common than in a non-heavy tailed distribution.

the systems to reach a normal steady state, and only then activate the page commit recording.

The user-level program that dumped the the ring buffer to disk—*the archival program*—was more complicated. We needed to ensure that it would not create page allocations itself, that it would successfully save most of the data from the ring buffer over a period of days, and that it would not create any secondary effects that would skew the data. To prevent dynamic page commits to this program, it was statically linked, employed no dynamic memory allocations, was started immediately upon boot (before the kernel instrumentation was enabled), and its pages were *wired*[2] with the `mlockall()` system call. These measures guaranteed that the archival program did not directly create memory allocations.

Indirect effects were possible since disk writes from the kernel instrumentation go through the buffer cache. A large write could result in cached data from other threads being flushed, causing unnecessary page allocations to those threads. We addressed this problem by limiting the amount of data that could be recorded in a single write to no more than five pages. The size of the buffer cache, hardcoded in the kernel, was approximately 5% of the total memory of a system, representing 6553 pages on the server (5% of 512 Mbytes, in 4096-byte pages) and 819 on the workstation (5% of 64 Mbytes). Our archival program therefore never wrote more than 0.08% of the server buffer or 0.6% of the workstation buffer. Finally, the scheduling of writes by the archival program had the potential to skew times of the page commits to other threads. We worked around this by having the writing process sleep for a Poisson period with a mean of 10 seconds between writes to disk. This did not change the fact that side-effects from archival program would occur, but since these effects would necessarily be correlated to the times at which it ran, this would simply inject a new set of Poisson arrivals to the system. Such a new arrival process would not alter the Poisson nature of the memory system, were it already to be Poisson.

The discussion above makes it clear that the design of the experiment was crucial, and many variations on the basic experiment are possible. For example, it is possible to send data either to a serial port or to the system console over a high-speed network connection, or even to completely bypass the operating system's disk buffer cache by performing synchronous writes. These alternatives all have the same disadvantage, viz. the incurred processor load becomes intolerable. A machine making 30,000 page commits in a second and writing all of them to a serial port would require a nearly one-megabit/second serial connection even if only three bytes were used to tag the commit. Assuming a 16-byte transmit buffer at the port and a "transmit buffer empty" interrupt at the completion of each buffer flush, such an experimental setup would generate over 5,500 processor interrupts each second. In reality, of course, many more bytes are required to accurately record the data we desired, and a similar load situation exists not only for serial port data, but also for console output or network communications. This data-serialization overhead and the associated processor interrupts significantly alter the processor utilization patterns: using any of these methods, the CPUs are either page faulting or else outputting page commit data during most time slices, and very little time is spent actually running user software. This is not effective experimental meter design. The same is true of synchronous writes to disk, since all system activity must be suspended while the physical disk platter spins to the desired position, the write is performed, and the write completes. In contrast, by using a minimal amount of system resources and taking advantage of the system's own caching, we were able to limit the space overhead to a known percentage of system memory, and to limit our experiment's influence on the time overhead to a controlled additive disturbance that would not skew our results.

[2]Wiring pages forces them to remain memory-resident. Once wired, the pages are never swapped or paged to secondary storage, but instead remain in semiconductor memory.

## III. Page Allocations Are Not Poisson

The difference between the CU-SERVER and the CU-WORKSTATION data reflects the influence that environment has upon memory usage. Figure 2 shows the complete time-series data from both machines at a one-second aggregate level. There are several interesting observations to be made about this data. First, bursts of arrivals in the page allocation data are much larger, relative to the background rate of the process, than those in the packet-trace data of Figure 1. In particular, there are a great many small page commits and only occasionally a huge one, so the arrivals have a notably different qualitative appearance from either Poisson or self-similar arrivals. Second, the roles of the machines are clearly apparent even in these overall activity plots: the server is generating page allocations almost continuously since it is always transferring email and answering DNS queries, whereas the workstation shows significant amounts of page allocation only during the weekdays, and then only during working hours— except for large bursts each night. This brings up another point: these are not isolated, self-sufficient machines, but rather part of a shared, dynamic network whose toolsets and parameters are under constant revision. In order to maintain consistency between the unix hosts on the network, configuration information is updated nightly through the remote file distribution service, `rdist`, a kind of remote file copy utility that can run executable commands before and after the copy. The scheduled execution of `rdist` causes a nightly burst of activity, which appears in both datasets.

An important observation that is not apparent from Figure 2 is that USERFIRST plus USERPOST allocations—that is, the first page allocation to a new nonroot-user thread combined with all other nonroot-user page allocations (see Table 1)—dominate the page allocation data in the CU-SERVER dataset, whereas SYSTEM allocations—kernel, root and nonfaulted allocations—dominate the CU-WORKSTATION dataset. Figures 3 and 4 show these categories, at the same one-second level of aggregation. This is not what we had anticipated. We expected to see 'system' commits dominating in the CU-SERVER dataset (taken from a machine experiencing little human-user activity) and ‘user’ commits dominating in the CU-WORKSTATION dataset (taken from a machine experiencing regular daily human activity). What appears to be workday activity exists in the SYSTEM allocations from the CU-WORKSTATION set; this is a result of help-desk personnel using root commands to do such things as changing user passwords, adding accounts, or editing general user information such as the `/etc/motd` file. Thus although most root threads are daemons such as `inetd` or `sendmail` and thus can be considered to be system-level activities (since their functions are largely automated), there exist *some* root-user threads that are normal human activities, making the separation between the two harder to make. In fact, the opposite is true as well, in that some threads that might be assumed to be human-initiated are actually automated system threads. Figure 4(a) shows that the CU-SERVER data is qualitatively identical to the ALL data in Figure 3, indicating that the overall arrival process is dominated by the USERFIRST and USERPOST page allocations. This is most likely a result of `named`’s domination of the system activity because of the enormous number of DNS queries and zone transfers to secondary nameservers that it satisfies. Unlike normal system daemons, the `named` daemon does not run as root but rather as the user ‘named,’ so its page allocations do not fall neatly into the SYSTEM category. We find, therefore, that in order to make a completely accurate definition of what constitutes a “user” or “system” activity, it is necessary to know beforehand (a) the specific environment in which the machine will operate, and (b) a complete breakdown of which daemons and commands fall into which category. Even then it is not clear how useful such a categorization might be; the `named` daemon, for example, might be changed in later releases to run as root and thereby change the categorized behavior. In short, categorization does not appear to be a good approach to characterizing memory commits.

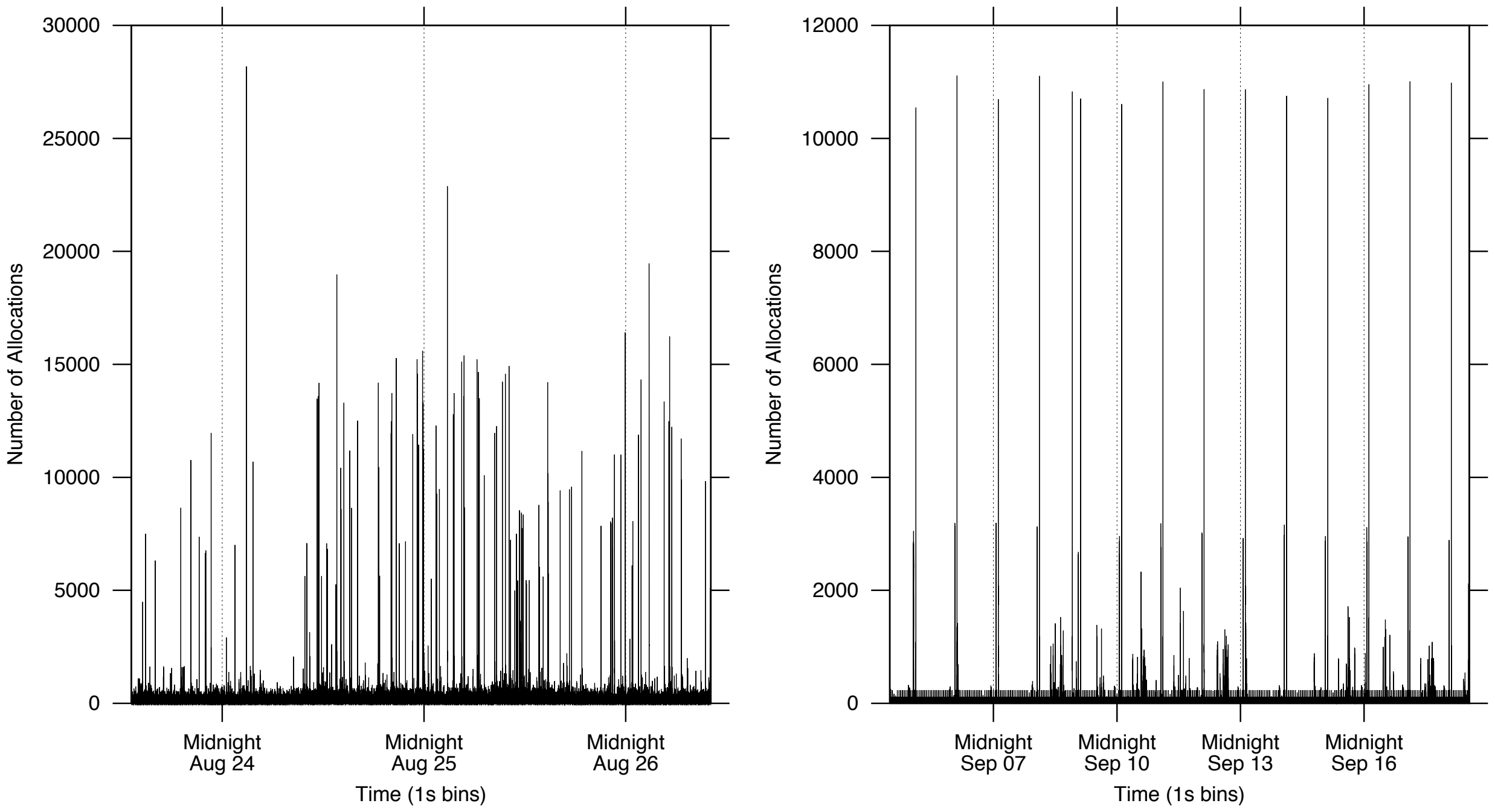

(a) CU-SERVER (b) CU-WORKSTATION

Fig. 2. ALL Page Allocations. All data from both machines, aggregated to a one-second level, with midnight represented by a dotted vertical line.

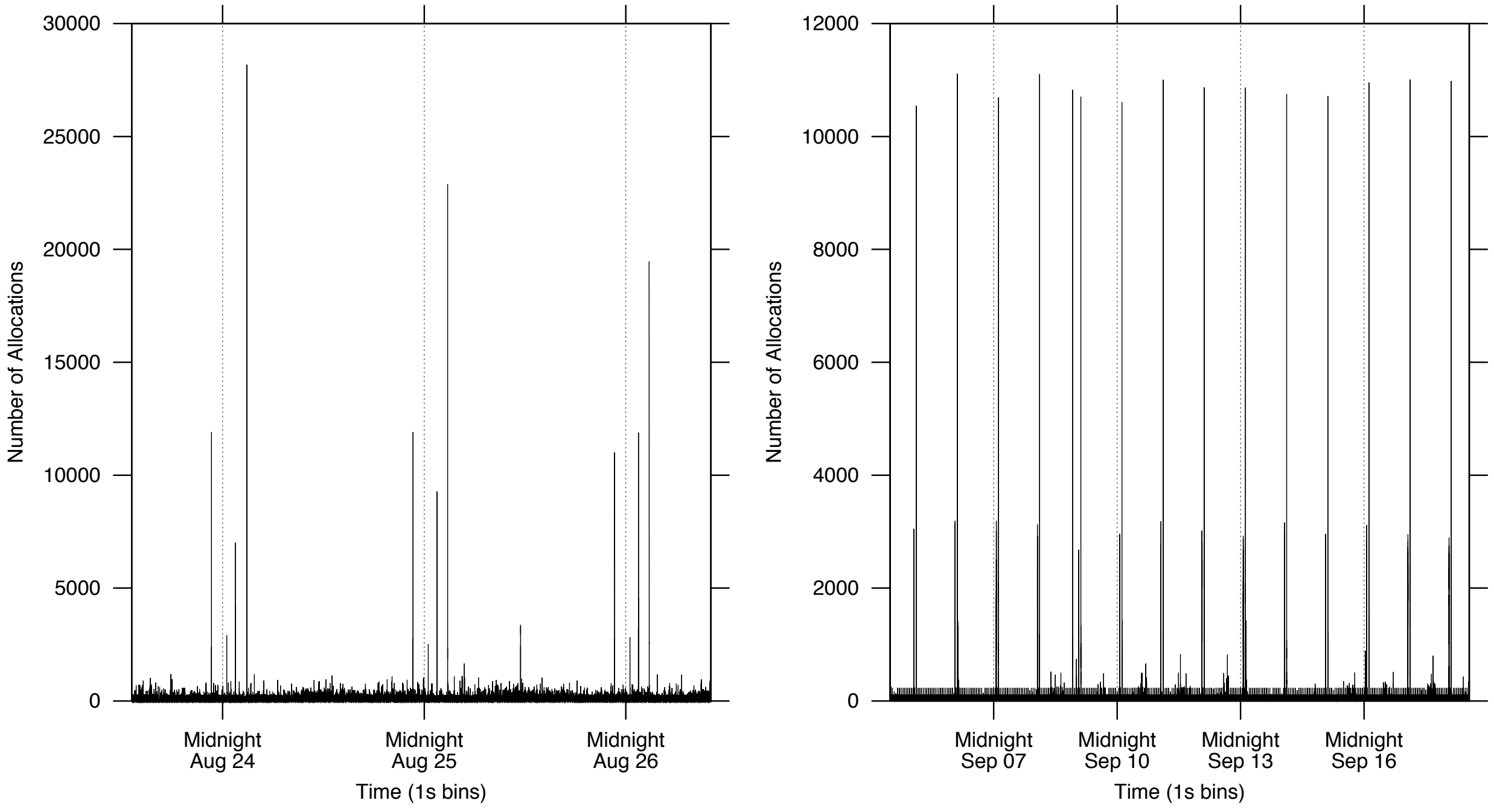

(a) CU-SERVER (b) CU-WORKSTATION

Fig. 3. SYSTEM Page Allocations. Both datasets display nightly, automatic system maintenance activity. The workstation data also displays system-like page commits during the daytime, as administrators engage in manual maintenance.

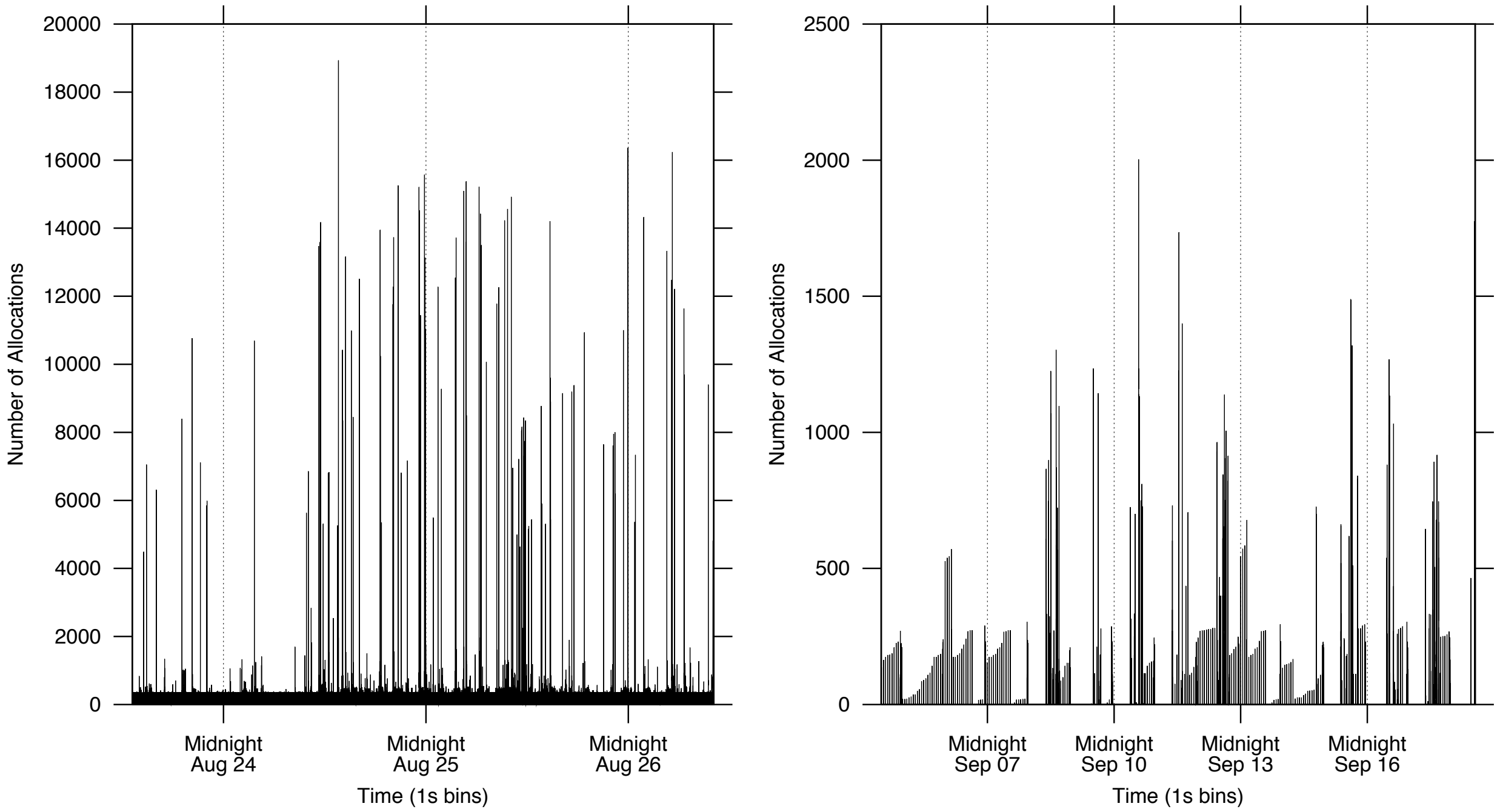

(a) CU-SERVER

(b) CU-WORKSTATION

Fig. 4. USERFIRST and USERPOST Page Allocations. Extensive user-like activity is present in the server dataset due to system threads running as nonroot users.

The data in the Figures above indicate that modeling by simple homogenous (fixed-rate) Poisson processes is not possible. Daily patterns exist in the data, and those patterns are different between the datasets. Certainly no homogenous Poisson process could accurately model all of these processes, since its mean arrival rate would need to be appropriate both for the quiescent, low-allocation periods present in both datasets *and* for their bursty, high-allocation periods. However, it might be possible to model short periods of individual datasets with separate Poisson processes. If so, then *piecewise* Poisson modeling (where different intervals are assigned different fixed rates) might be an effective technique.

To investigate piecewise Poisson modeling of the page commit arrival process, we examined both datasets using a statistical goodness-of-fit test, comparing the arrivals over various sub-intervals against the expected arrivals from a Poisson process. We chose intervals ranging from one second to over nine hours as representative of those over which software controllers might be expected to operate, but did not calculate the statistic for intervals less than one second due to the sparseness of the sample data. Our test was the reduced-$\chi^2$ method, a straightforward method (described in many texts, e.g. [4]) that associates significance levels with results in order to determine the closeness of fit. Given a random variable and its hypothesized distribution, the reduced-$\chi^2$ method allows one to calculate the difference between the values of that variable and the expected values from the distribution, as well as an associated significance level. For example, if the value from a reduced-$\chi^2$ test calculation is associated with a significance level of 90% it means that with probability 0.9, the sample data comes from a random variable with the same distribution as the hypothesized distribution (the null hypothesis). If the sample data agrees with the null hypothesis to a significance level of 95%, the agreement is said to be *significant*; if it fits to within 99%, it is called *highly significant*. Results below these levels are assumed to not confirm the null hypothesis.

We calculated the reduced-$\chi^2$ statistic for both datasets in four different categories (ALL, SYSTEM, USERFIRST, USERFIRST+USERPOST), studying different timescales ranging from one second to approximately nine hours. Figure 5 shows the results up to intervals of about 10 minutes, at which point the data had converged. The *highly significant* and *significant* levels are indicated by horizontal lines. That is, any data points on the graphs of Figure 5 that are above the 95% or *significant* line can be said to be consistent with Poisson arrivals; all others are not. It is apparent that most of the intervals do not meet the the significant or highly-significant level. In general—i.e., all allocations combined—page commits do not fit a Poisson arrival distribution at *any* timescale. In fact only the USERFIRST data—and only when viewed over extremely small intervals—demonstrates Poisson arrivals with statistical significance. Interestingly, the USERFIRST arrivals in the CU-SERVER dataset tend more towards Poisson arrivals than the USERFIRST arrivals in the CU-WORKSTATION dataset. Since `named` dominates this data, this implies that DNS queries are arriving according to a Poisson process, which makes sense: DNS queries are largely independent, they are often initiated by human activity (e.g. logins to machines, or web browsing), they are initiated randomly (i.e., with exponential interarrival times), and they are numerous since almost all network communication begins with such a query.

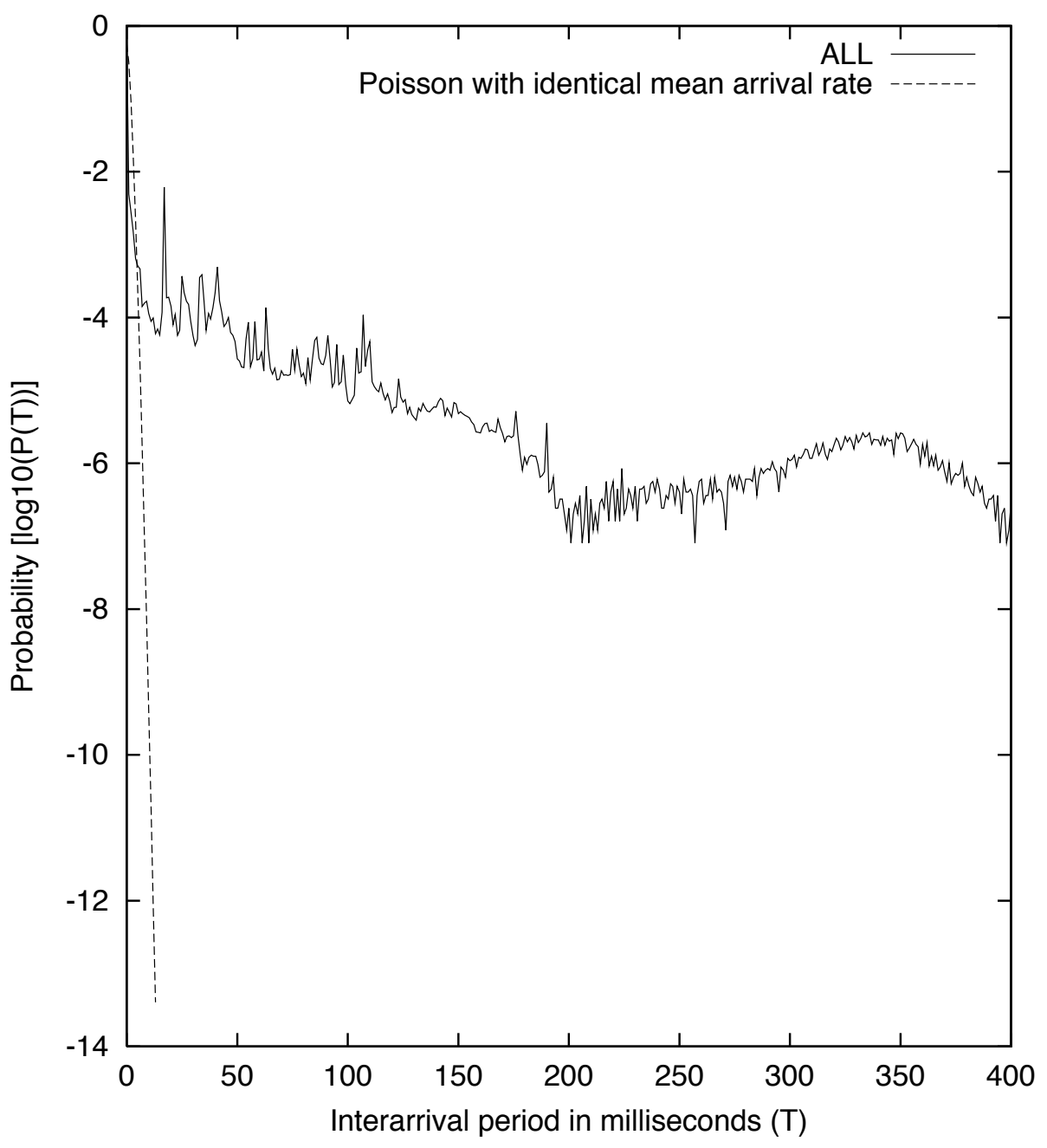

(a) CU-SERVER

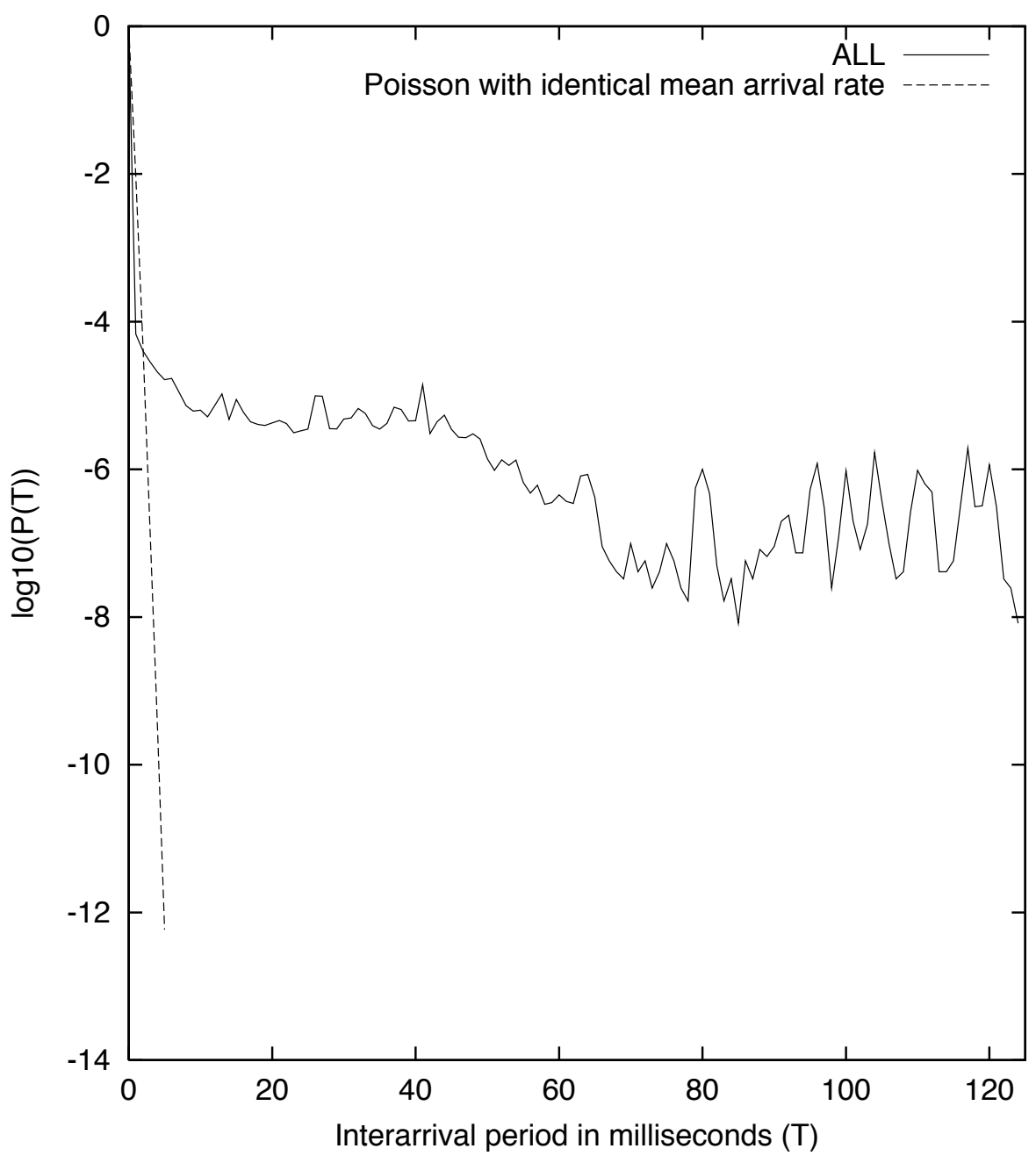

(b) CU-WORKSTATION

Fig. 5. Results of modeling arrivals with a piecewise Poisson process. Each data point represents the reduced-$\chi^2$ significance level for a single fixed interval. Data points above the 95% level are consistent with Poisson arrivals.

These statistical tests demonstrate that the data is clearly not consistent with Poisson models. We next investigate whether the memory arrival process might be self-similar.

## IV. Page Allocations Are Probably Not Self-similar

Heavy-tailed distributions and persistent high variance, which indicate burstiness over long timescales, have come to be associated with *self-similar* stochastic processes [25]. The heavy tails and high variance in self-similar processes are the result of highly correlated, aperiodic bursts of activity. We now show that unix memory allocations from both of our study datasets are bursty and possess persistent variance, but are most likely not self-similar.

A *heavy tailed* distribution is defined as one that has a cumulative distribution function that decays more slowly than an exponential [8]. An exponential cumulative distribution function (cdf) is defined as $F(x) = 1 - e^{-x/\mu}$, where $\mu$ is the mean. Typically, one demonstrates heavy tails by plotting $x$ versus $1 - F(x)$ on a semilog graph, where an exponential cdf appears as a straight line with a negative slope. Any distribution function above this line is then called heavy tailed. This can also be seen by plotting the probability distribution functions; a heavy-tailed process (as suggested by its name) will exhibit more positive probability mass in the tail than an exponential one with the same mean.

For the purposes of this paper, we take a similar approach and call our data heavy tailed if the empirical probability distribution functions (pdf) of the allocation data decay more slowly than the pdf of a Poisson process with the same mean as the tested data[3]. The empirical pdfs for the ALL category of data on both systems are shown in Figure 6; the remaining categories are qualitatively identical. The pdf for a Poisson process with the same mean as the respective ALL dataset is drawn in each figure, appearing as a dashed line with a negative slope. The empirical pdfs clearly decay much more slowly than their Poisson counterparts: both have extremely heavy tails. This implies that the probability of the arrival of a new request for memory does not decrease exponentially with time, as it would with a Poisson process—i.e., that relatively long periods of quiescence are likely. Therein lies a clue to the basis of the relationship between heavy tails and "burstiness:" a burst is a period of quiet followed by a set of arrivals, so the likelihood of bursts increases along with the likelihood of quiescent periods. The existence of heavy tails is therefore an important consideration in controller design: a control system that works with Poisson-distributed arrivals need not be able to react to changes as quickly as one that handles arrival distributions with heavy tails, as the latter may demand rapid changes in service while the former will not.

To show that the memory allocation arrival process not only is heavy tailed, but also possesses persistent high variance, we compare the different allocations against Poisson distributions using the time-variance technique [14], [19]. Following this method, we first examine the data at a fine resolution—perhaps 0.01 seconds—and calculate the variance of the data at that level. We then smooth the data by aggregating the first ten 0.01-second bins into a single bin, whose number of arrivals is equal to the mean of the 10 smaller bins, and so on, proceeding in this fashion until the entire dataset is aggregated to the one-second level. We then calculate the variance of the new dataset, which should be expected to be lower than the original value, since aggregation smooths bursts.

[3]The pdf is the derivative of the cdf, when that derivative exists. If $F(x)$ is a cdf, then $F(5)$ represents the probability that $x \leq 5$; $F'(5)$ would then be the probability that $x = 5$. The empirical pdf is the estimated value of the pdf of an arrival process based on its history.

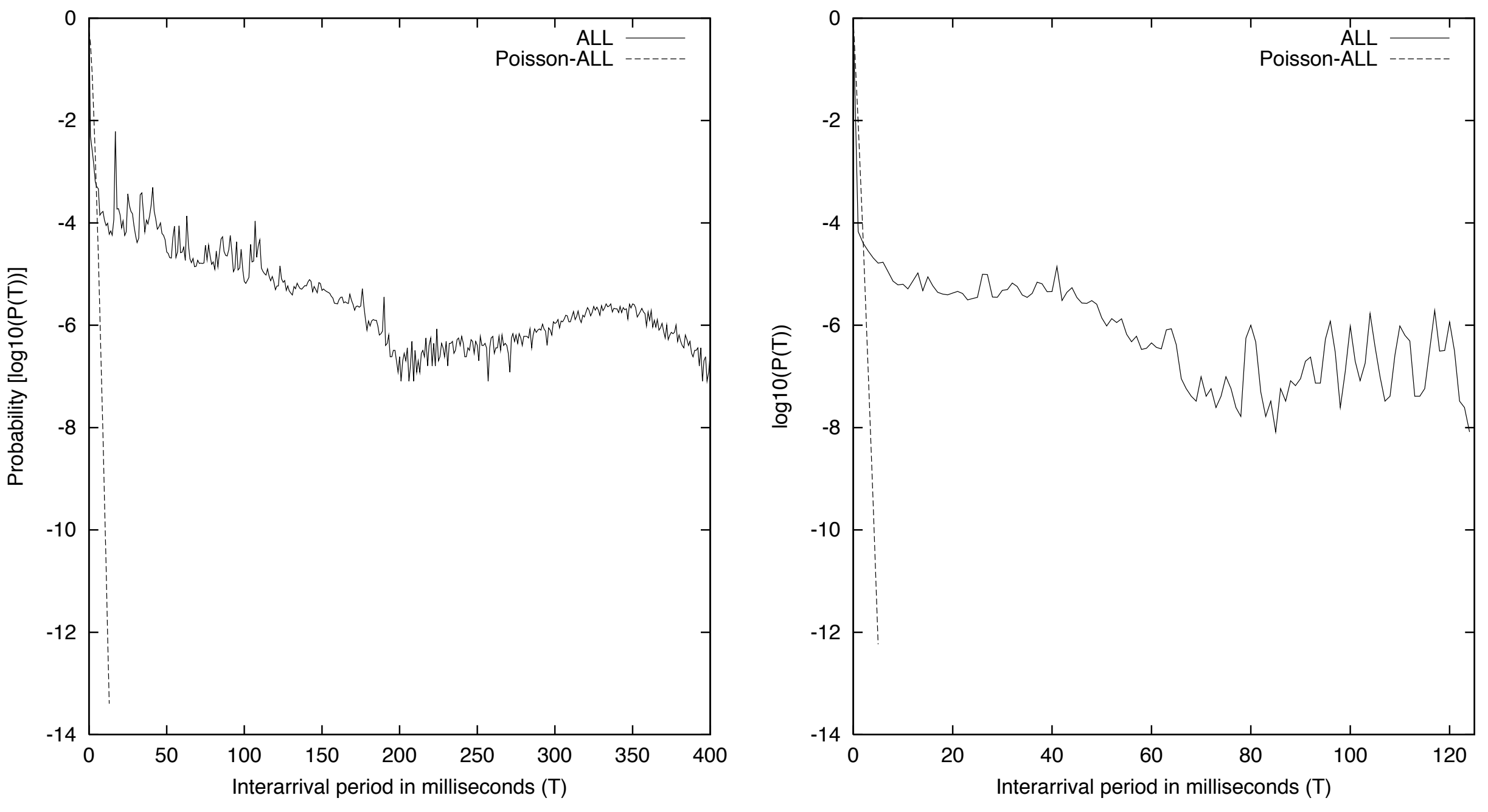

(a) CU-SERVER

(b) CU-WORKSTATION

Fig. 6. Empirical probability distribution functions for the ALL process from both machines, compared to Poisson processes with the same respective means. The empirical pdfs decay more slowly than the Poisson processes, making them heavy-tailed.

For example, given two bins containing $10$ and $0$ arrivals, respectively, the mean is $5$, and the variance is $(5^2+5^2)/2 = 25$. If we combine the two bins into one, the new mean is still $5$, but the variance has been reduced to $0$, meaning that no bursts remain. A Poisson process exhibits a characteristic slope on a time-variance plot: since its variance decays as $1/\epsilon$ at aggregation level $\epsilon$, a time-variance graph of Poisson data (which plots the log of the aggregation level versus the log of the variance) is a straight line with slope $= -1$.

The time-variance analysis technique can reveal not only Poisson behavior, but also self-similarity. Slopes on time-variance curves that approach a limiting value (not equal to $-1$) indicate a power-law scaling phenomenon at work, a characteristic of self-similarity [16], [20]. If such a limiting value exists then it is possible to estimate the degree of self-similarity, a value known as the *Hurst* parameter and denoted by $H$ [19]. When such an estimate of $H$ can be made, it is possible to describe the process as being self-similar.

To determine if the variance of our page commit data was consistent with Poisson or self-similar scaling, we plot the time-variance characteristics of the ALL, SYSTEM, USERFIRST plus USERPOST, and USERFIRST arrival processes, as shown in Figures 7 and 8. The initial value of the variance of each at an aggregation level of $0.01$ seconds was used for normalization, so each curve originates at $y = 0$. In addition, we have drawn a line from a variance ($y$-axis) of zero with a slope of -1; if the data were consistent with a Poisson process, it would roughly fit this line. These figures show that, in general, the memory allocation arrival process does *not* have a Poisson variance over a wide range of timescales— although the CU-SERVER USERFIRST curve in Fig 8(c) shows a slope approaching -1 at timescales less than approximately $10^2$ seconds, which is consistent with the reduced-$\chi^2$ analysis of Section 3.

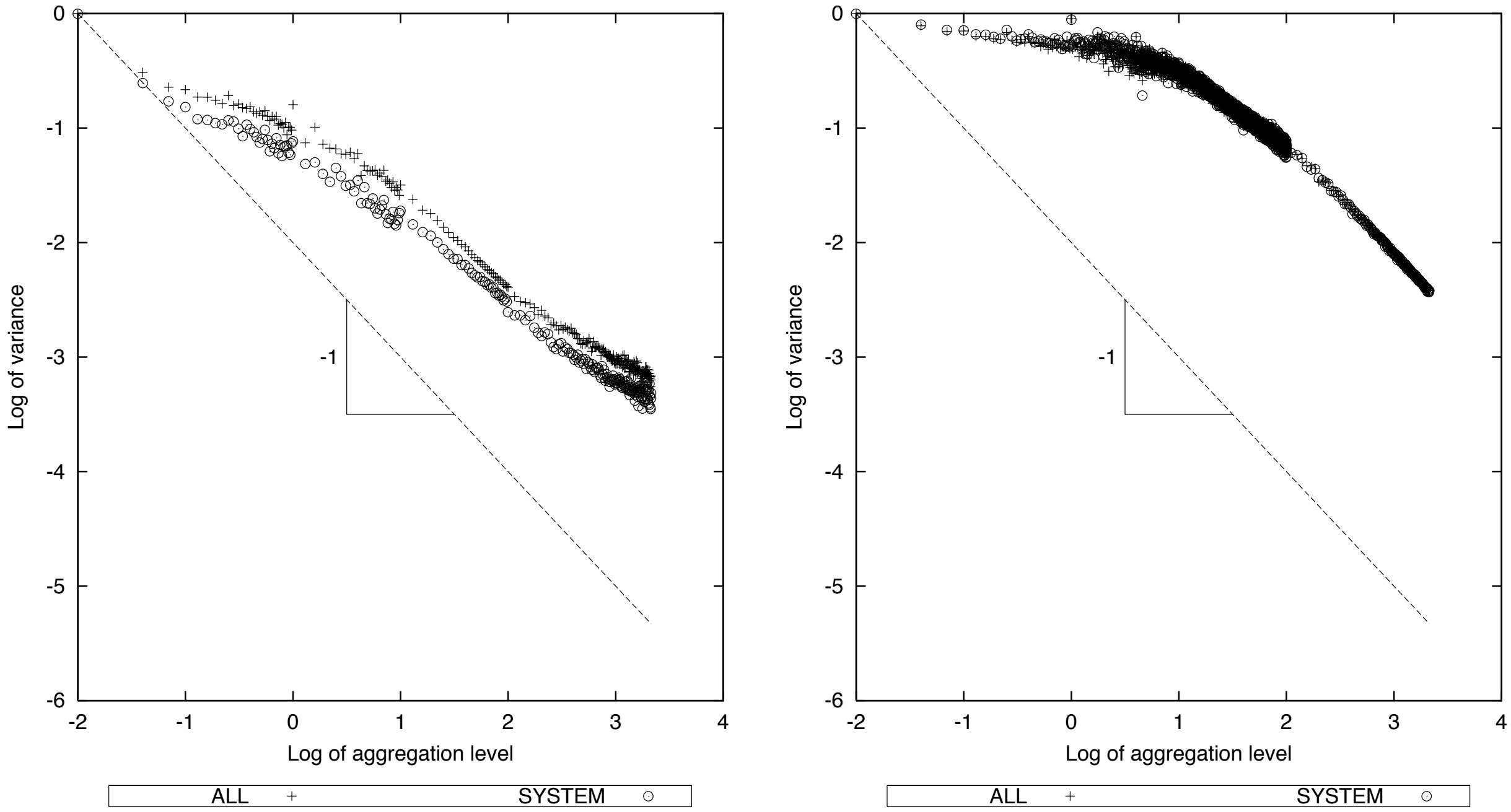

(a) CU-SERVER

(b) CU-WORKSTATION

Fig. 7. Time-variance plot of system allocations. If the data possesses a slope of $-1$, it is consistent with a Poisson process; if it possesss any other fixed slope, it is consistent with a self-similar process.

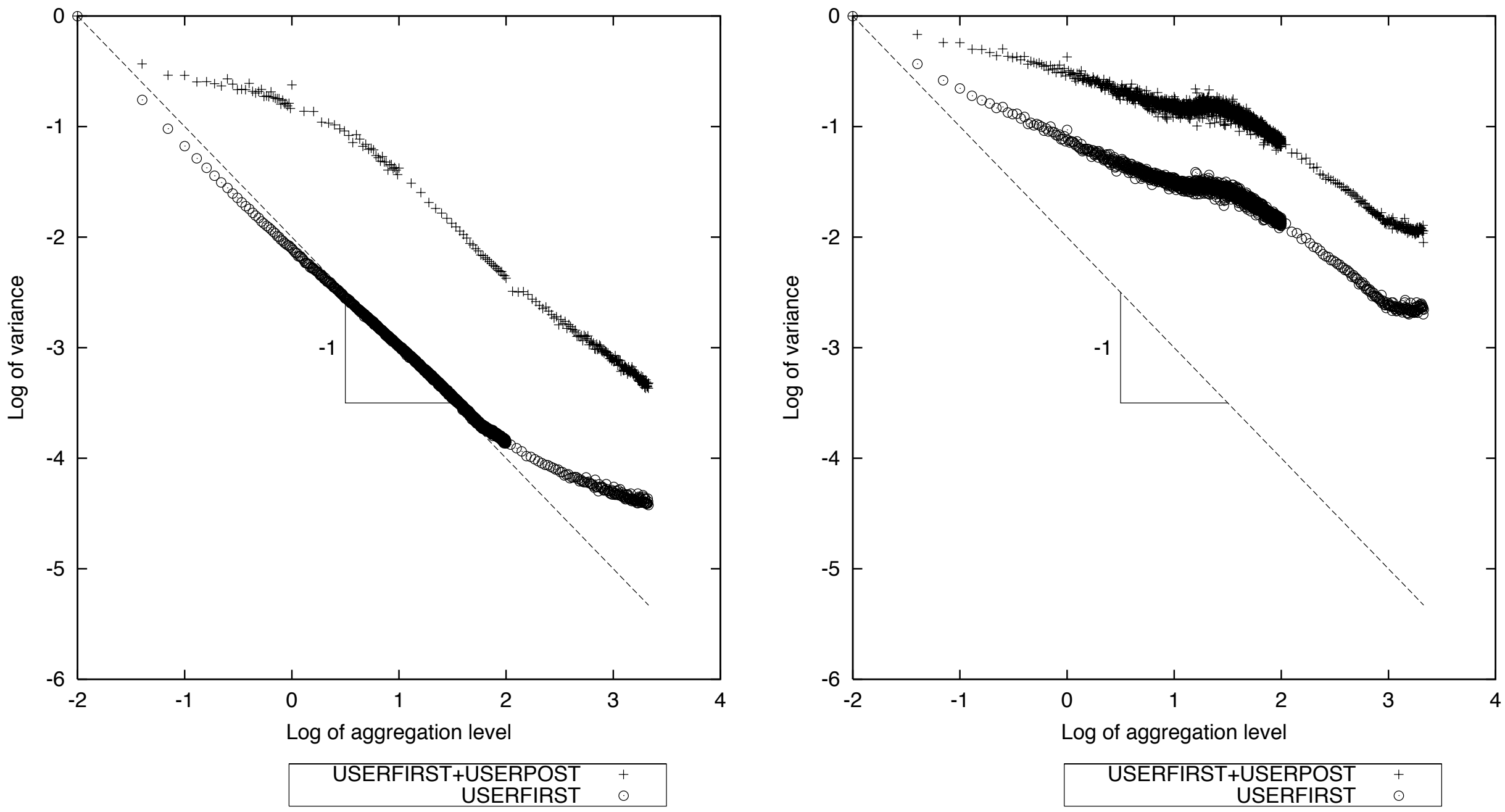

(a) CU-SERVER

(b) CU-WORKSTATION

Fig. 8. Time-variance plot of user allocations.

The results in Figures 7 and 8 indicate that the memory allocation process exhibits non-Poisson variance that persists over long timescales. This is most likely the result of the extremely large allocation bursts that we found in the data—bursts that require many iterations of aggregation and smoothing before converging to the overall mean. Because of this, a single estimate of the Hurst parameter $H$ is not possible since no limiting slopes are approached by the processes shown in Figures 7 or 8. When such limits do appear to be present, as in the USERFIRST+USERPOST data of Figure 8 at log aggregations above 1, the slope is consistent with *Poisson* processes rather than self-similar ones. Because no single estimate of $H$ can be made, the data indicate that page commits are not strictly self-similar at all timescales, although it is perhaps justified to conjecture that self-similar models might be appropriate over small intervals for some kinds of allocations. This implies that heavy tails and persistent variance are not reliable indicators of self-similarity.

## V. No Simple Distribution Suffices

From the same data that we used to show what the memory commit process *is not*, we can infer useful clues about what it *is*. These clues do not extend as far as allowing us to precisely characterize the arrival process with only the (rather rough) categorizations of kernel, root, userfirst, userpost and nonfault commits, but they are enough to draw the conclusion that *no* simple process or distribution can accurately capture the dynamics of the commit process. This means that no neat modeling assumptions can be safely made about the memory commit process, and that the attendant tricks of tuning controllers for specific arrival distributions that have been used for network traffic (e.g. [11], [12]) are not applicable to memory commits.

A hint of the dynamics of the commit process can be found in the distribution of delays between page commits, i.e., how many times a particular-length calm period occurred. We examined all 31 combinations of the categories in the SERVER data and plotted the distribution of delays therein. We found two dominant kinds of behavior in these distributions, depending upon the categories of commits under consideration. Representative examples are shown in Figure 9; the remaining 29 combinations of categories were qualitatively similar to one of these other plots. The data in both subfigures is aggregated into 1000$\mu$s bins, the finest resolution of the raw commit data. The distribution of delays in Figure 9(a) represents ALL commits, i.e., the complete dataset. Notable features in this data are a pair of roughly linearly-decaying, interleaved traces marked ‘Upper’ and ‘Lower’, a surge of commits all occurring after a roughly 3000ms-long delay marked ‘Anomaly’, and a sharp break in the upper trace marked ‘Dropoff’. Similar linear decays seem to be present in Figure 9(b) (the USERPOST data), but it is hard to tell.

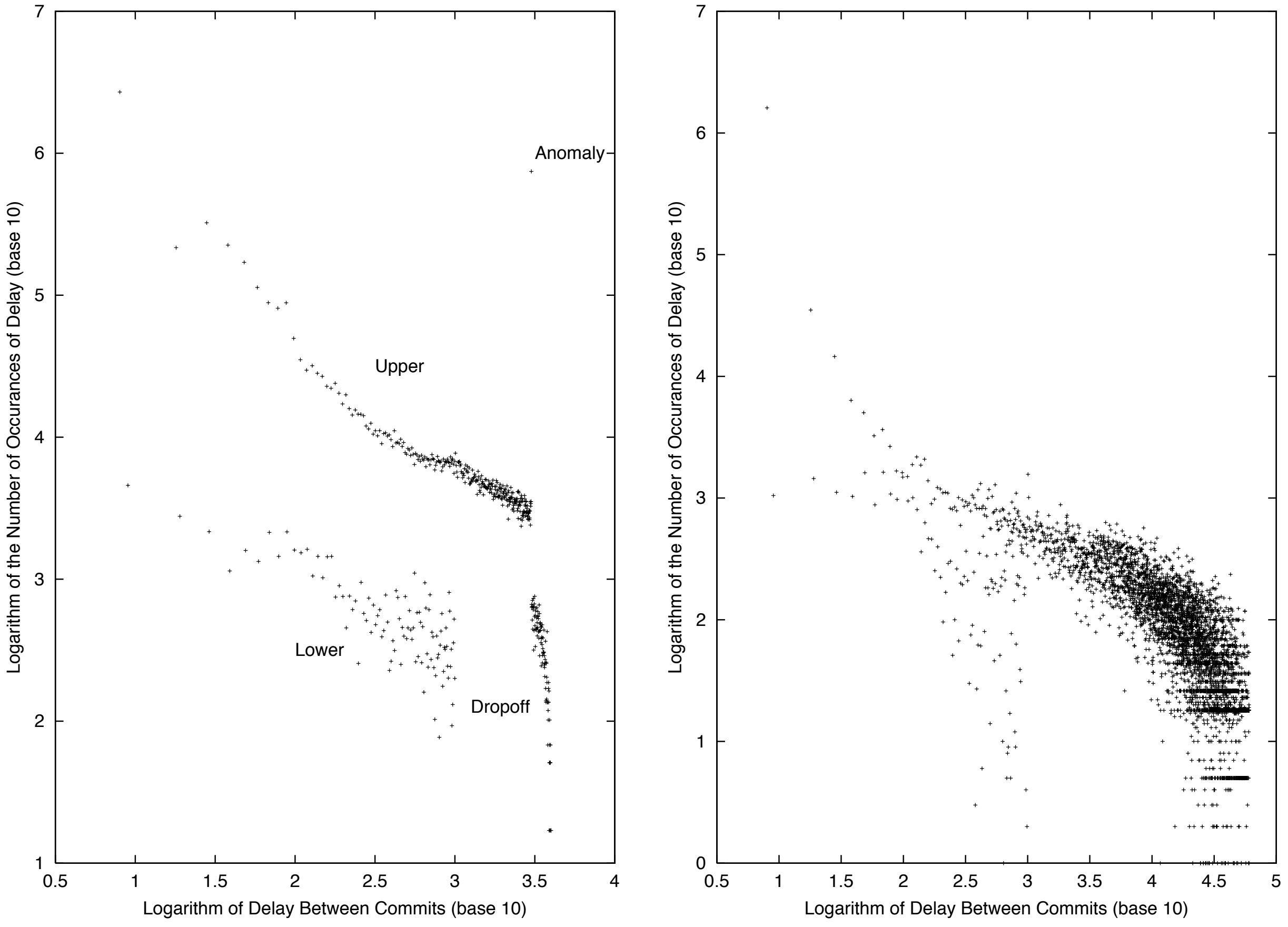

(a) ALL Commits

(b) USERPOST Commits

Fig. 9. Distribution of delays between SERVER page commits (1000$\mu$s bins). Structure consists of two interleaved arrival processes (Upper and Lower), an anomalous point (Anomaly) and a tail (Dropoff) in subfigure (a), and a roughly linear arrival process in subfigure (b). Quantization in the lower righthand corner of subfigure (b) is an artifact of the log/log plotting process.

Since our experiment did not record the threads or kernel subsystems that obtained memory pages, it is impossible for us to conclusively state what operating system activities created the features in Figure 9, but we can conjecture about some causes and relationships. Linear regression can be used to fit lines to the data in just the Upper and Lower traces, with slopes of $-0.90$ and $-0.65$, respectively; the similarity in these slopes (when compared to the other traces) suggests that the traces are correlated to the same underlying process. It is uncertain what this process might be, but the features of Figure 9(a) are present in all SERVER data combinations that include either ROOT or NONFAULT page commits, indicating that a user-level root process is responsible. One such process on the server machine is the `sendmail` thread. The data point labelled 'Anomaly' is a surge of page commits occurring after a quiet period of slightly more than 3000ms. This surge appears in Figure 9(a) at $x \approx 3.5$ ($\log_{10} 3000 = 3.47$). This period of approximately three seconds is strongly suggestive of the initial timeout delay associated with TCP/IP packet retransmissions [29], which, according to IETF recommendations, should be three seconds [24]. A likely mechanism for the surge of commits, then, is a sendmail thread receiving data from a peer over a busy network. On such a network, the thread would be causing a steady flow of page commits with three-second pauses whenever transfers from the peer are corrupted, retransmitted, and then successfully completed. Since the SERVER machine in our data was the department mail hub, this scenario is reasonable. The distinct break in the Upper trace after the anomaly (the 'Dropoff') would then imply that delays greater than the TCP/IP initial retransmission timeout between page commits are extremely unlikely, although any attempt at explaining why using the present data would be highly speculative. Overall, it is clear from the complex structure of the data that a single probablity distribution function would be insufficient for modeling the SERVER data arrival process.

If any single distribution *would* suffice, at least in part, it would be one that results in linear traces on log/log axes like those in Figure 9. Linearity on such axes is indicative of of power law relationships of the form $y = Cx^{-a}$, which can be modeled with a power law distribution such as the Zipf or— more commonly—the Pareto distribution, whose probability distribution function (PDF) is given by $p_X(x) = km^k x^{-(k+1)}$, where $m > 0$, $k > 0$, $k \geq m$ and $k, m$ are constant parameters that define the specific shape of the distribution. When plotted on log/log axes, this PDF takes the form $\log(p_X(x)) = \log k + k \log m - (k+1) \log x$, a linear relationship. For the SERVER data, then, the interleaved linear processes of Figure 9(a) could be said to be two Pareto arrival processes up to the 3-second limit imposed by the TCP/IP subsystem.

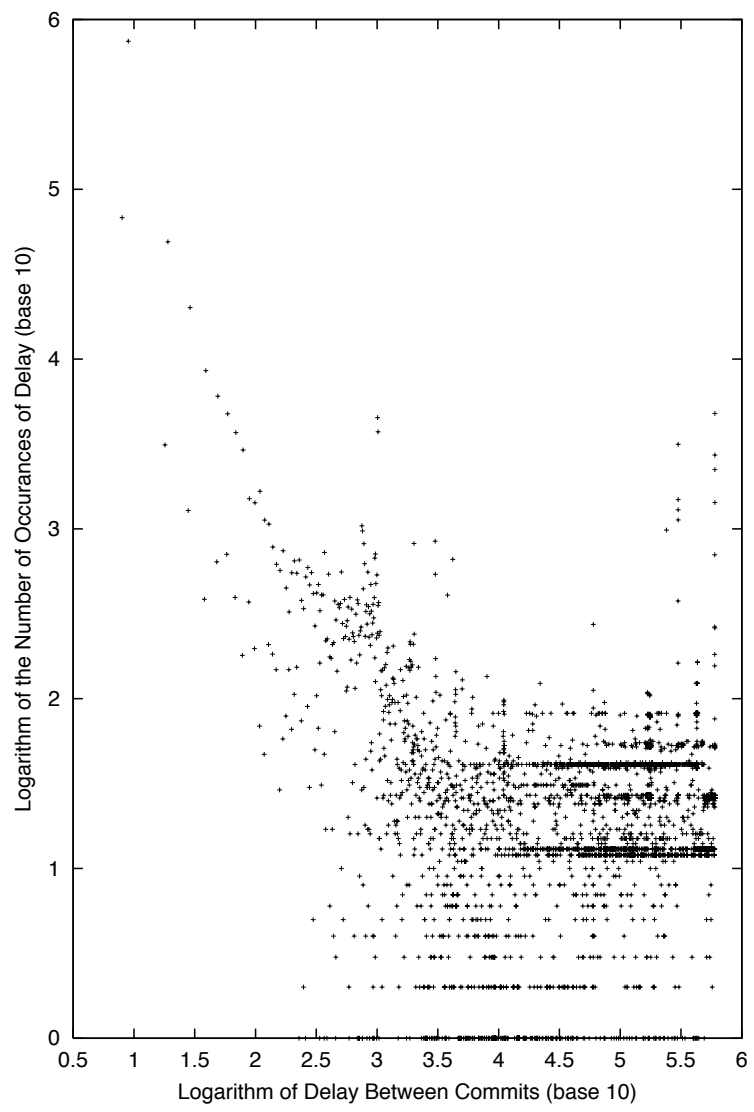

Fig. 10. Distribution of Delays Between WORKSTATION ALL Page Commits (1000$\mu$s bins). Very little structure appears to exist.

Unlike the SERVER data, the WORKSTATION data shows little structure of any kind (other than quantization effects introduced by the plotting) and does not appear to be amenable to modeling, even with Pareto distributions. An examination of the 31 combinations of categories for this dataset showed all to be qualitatively similar. A representative example is the ALL dataset, shown in Figure 10.

In summary, it is apparent that no single dis-tribution suffices to model both the SERVER and WORKSTATION data, or even just the SERVER data. The structure in the SERVER data appears to be the result of the threads that the machine was running and its networked environment, so that modeling its memory commit behavior would be, in effect, modeling those threads' memory requests and that environment.

## VI. Conclusion

We investigated two classes of models that have been applied to memory allocation and network packet arrivals. In particular, we have shown experimentally that page commits under UVM in OpenBSD-2.9 do not follow a Poisson distribution, but instead possess persistent high variance over long timescales. Our experiments involved production unix machines in very different working environments. Neither servers nor workstations were Poisson: except in one specific case (first-page allocations by user processes, analyzed over short time intervals), the arrivals were not exponentially distributed. Since most general-purpose unix hosts operate in one of these two roles (or some combination), this study is widely representative.

These results are important for any model of virtual memory systems that uses standard queueing theory techniques—methods that usually assume Poisson arrivals. The errors introduced by this assumption depend on the variance of the allocation process, whose extremely bursty nature skews the average rate of the arrivals towards high values that do not accurately represent instantaneous arrivals rates. Models that assume a particular mean arrival rate will therefore predict more memory usage than necessary for most purposes, and not enough in others. This is a particularly important result for dynamic, real-time resource control systems, as it implies that modeling and control of systems that are operating near their resource limits requires theory that incorporates the burstiness of the allocation process. The currently accepted models do not do this.

We have also shown that memory allocations are not well described by stationary self-similar models, another widely used formalism in stochastic traffic modeling, although more research is required to confirm or refute this hypothesis. Our time-variance analyses suggest that nonstationary self-similar models might be effective at some timescales, but we believe that the close relationship between the tasks performed by machines in different working environments and the resulting page allocation process will make this difficult to demonstrate in general. This conclusion is supported by our analysis of the temporal distribution of page commits on both machines, which indicates that the working environment can create (or fail to create) structure in the distribution. The presence of data that fit a line on log/log plots of these distributions hints at the possibility of using Pareto models for memory commits, but more detailed work is necessary in order to evaluate the effectiveness of any such scheme.

In the future we plan to itemize the exact threads that use memory and also track the page *release* process. In the first case, knowledge of what threads use memory would make it possible to determine what threads are correlated, and how these correlations are reflected in the structure of the page commit distributions. In the second case, maintaining a record of page releases would create an exact record of memory usage that could be used to calculate an *a posteriori* record of the size of the available page pool, along with empirical distributions over different periods of time, which may prove useful to the design of controllers for virtual memory systems. Although it is clear that Unix memory allocations are not Poisson, better characterizations of the commit arrival process should be possible using the same methods described in this paper.

PLACE PHOTO HERE

**James Garnett** (insert biography text)

PLACE PHOTO HERE

**Elizabeth Bradley** (insert biography text)